\documentclass[aps,preprint]{revtex4}%
\usepackage{amsfonts}
\usepackage{amsmath}
\usepackage{amssymb}
\usepackage{graphicx}%
\providecommand{\U}[1]{\protect\rule{.1in}{.1in}}

\begin{document}
\preprint{ }
\title{Inter-band magnetoplasmons in mono- and bi-layer graphene}
\author{M. Tahir$^{1\ast}$}
\affiliation{Department of Physics, University of Sargodha, Sargodha, Pakistan.}
\author{K. Sabeeh$^{2}$}
\affiliation{Department of Physics, Quaid-i-Azam University, Islamabad, Pakistan.}
\pacs{PACS number}

\begin{abstract}
Collective excitations spectrum of Dirac electrons in mono and bilayer
graphene in the presence of a uniform magnetic field is investigated.
Analytical results for inter-Landau band plasmon spectrum within the
self-consistent-field approach are obtained. SdH-type oscillations that are a
monotonic function of the magnetic field are observed in the plasmon spectrum
of both mono- and bi-layer graphene systems. The results presented are also
compared with those obtained in conventional 2DEG. The chiral nature of the
quasiparticles in \ mono and bilayer graphene systems results in the
observation of $\pi$ and $2\pi$ Berry's phase in the SdH- type oscillations in
the plasmon spectrum.

\end{abstract}
\maketitle

\section{I. Introduction}

Recent progress in the experimental realization of both monolayer and bilayer
graphene has led to extensive exploration of the electronic properties in
these systems\cite{1,2}. Experimental and theoretical studies have shown that
the nature of quasiparticles in these two-dimensional systems is very
different from those of the conventional two-dimensional electron gas (2DEG)
systems realized in semiconductor heterostructures. Graphene has a honeycomb
lattice of carbon atoms. The quasiparticles in monolayer graphene have a band
structure in which electron and hole bands touch at two points in the
Brillouin zone. At these Dirac points the quasiparticles obey the massless
Dirac equation leading to a linear dispersion relation $\epsilon_{k}=v_{F}k$
(with the Fermi speed $v_{F}=10^{6}m/s)$. This difference in the nature of the
quasiparticles in monolayer graphene from conventional 2DEG has given rise to
a host of new and unusual phenomena such as the anamolous quantum Hall effects
and a $\pi$ Berry phase\cite{1,2}. These transport experiments have shown
results in agreement with the presence of Dirac fermions. The 2D Dirac-like
spectrum was confirmed recently by cyclotron resonance measurements and also
by angle resolved photoelectron spectroscopy (ARPES) measurements in monolayer
graphene\cite{3}. Recent theoretical work on graphene multilayers has also
shown the existence of Dirac electrons with a linear energy spectrum in
monolayer graphene\cite{4}. On the other hand, experimental and theoretical
results have shown that quasiparticles in bilayer graphene exhibit a parabolic
dispersion relation and they can not be treated as massless but have a finite
mass. In addition, The quasiparticles in both the graphene systems are
chiral\cite{2,4,5,6,7}.

Plasmons are a very general phenomena and have been studied extensively in a
wide variety of systems including ionized gases, simple metals and
semiconductor 2DEG systems. In a 2DEG, these collective excitations are
induced by the electron-electron interactions. Collective excitations
(plasmons) are among the most important electronic properties of a system. In
the presence of an external magnetic field, these collective excitations are
known as magnetoplasmons. Magnetic oscillations of the plasmon frequency occur
in a magnetic field. Single particle magneto-oscillatory phenomena such as the
Shubnikov-de Haas and de Haas-van Alphen effects have provided very important
probes of the electronic structure of solids. Their collective analog yields
important insights into collective phenomena \cite{8,9,10,11,12,13,14,15}.
Collective excitations of Dirac electrons in monolayer and bilayer graphene in
the absence of a magnetic field have been investigated \cite{16,17,18,19,20} .
Magnetic field effects on the plasmon spectrum have not been studied so far.
In addition, since the quasiparticles in graphene are chiral, the particles
will acquire Berry's phase as they move in the magnetic field leading to
observable effects on the plasmon spectrum. To this end, in the present work,
we study the magnetoplasmon spectrum within the self-consistent-field approach
for both the monolayer and bilayer graphene systems. Magnetoplasmons can be
observed by in-elastic light scattering experiments as revealed in studies
carried out on 2DEG systems \cite{11,12,13,14,15}. Similarly, in-elastic light
scattering experiments are expected to yield information about the
magnetplasmons in graphene. Furthermore, the results presented here can also
be experimentally observed by Electron Energy Loss Spectroscopy (EELS) on
graphene \cite{21}.

\section{Electron energy spectrum in monolayer graphene}

We consider Dirac electrons in graphene moving in the $x-y$-plane. The
magnetic field ($B$) is applied along the z-direction perpendicular to the
graphene plane. We employ the Landau gauge and write the vector potential as
$A=(0,Bx,0)$. The two-dimensional Dirac like Hamiltonian for single electron
in the Landau gauge is ($\hbar=c=1$ here) \cite{1,2}%
\begin{equation}
H_{0}=v_{F}\sigma.(-i\nabla+eA). \label{1}%
\end{equation}
Here \ $\sigma=\{\sigma_{x},\sigma_{y}\}$are the Pauli matrices and $v_{F}$
characterizes the electron Fermi velocity. The energy eigenfunctions are given
by%
\begin{equation}
\Psi_{n,k_{y}}(r)=\frac{e^{ik_{y}y}}{\sqrt{2L_{y}l}}\left(
\begin{array}
[c]{c}%
-i\Phi_{n-1}[(x+x_{0})/l]\\
\Phi_{n}[(x+x_{0})/l]
\end{array}
\right)  \label{2}%
\end{equation}
where%
\[
\Phi_{n}(x)=\frac{e^{-x^{2}/2}}{\sqrt{2^{n}n!\sqrt{\pi}}}H_{n}(x),
\]
$l=\sqrt{1/eB}$ is the magnetic length, $x_{0}=l^{2}k_{y},$ $L_{y}$ is the
$y$-dimension of the graphene layer and $H_{n}(x)$ are the Hermite
polynomials. The energy eigenvalues are%
\begin{equation}
\varepsilon(n)=\omega_{g}\sqrt{n} \label{3}%
\end{equation}
where $\omega_{g}=v\sqrt{2eB}$ is the cyclotron frequency of the monolayer
graphene and $n$ is an integer$.$ Note that the Landau level spectrum for
Dirac electrons is significantly different from the spectrum for electrons in
conventional 2DEG\ which is given as $\varepsilon(n)=\hbar\omega_{c}(n+1/2)$.
The Landau level spectrum in graphene has $\sqrt{n}$ dependence on the Landau
level index as against linear dependence in 2DEG. The monolayer graphene has
four fold degenerate (spin and valley) states with the $n=0$ level having
energy $\varepsilon(n=0)=0.$ The quasiparticles in this system are chiral
exhibiting $\pi$ Berry's phase.

\section{Electron energy spectrum for bilayer graphene}

The Landau level energy eigenvalues and eigenfunctions are given by\cite{5}%
\begin{equation}
\varepsilon(n)=\omega_{b}\sqrt{n(n-1)}, \label{4}%
\end{equation}%
\begin{equation}
\Psi_{n,K}^{\pm}=\frac{1}{\sqrt{2}}\left(
\begin{array}
[c]{c}%
\Phi_{n}\\
\pm\Phi_{n-2}\\
0\\
0
\end{array}
\right)  , \label{5}%
\end{equation}%
\begin{equation}
\Psi_{n,K^{\prime}}^{\pm}=\frac{1}{\sqrt{2}}\left(
\begin{array}
[c]{c}%
0\\
0\\
\pm\Phi_{n-2}\\
\Phi_{n}%
\end{array}
\right)  , \label{6}%
\end{equation}
where $\pm$\ assigned to electron and hole states, $\omega_{b}=\frac
{eB}{m^{\ast}}$ is the cyclotron frequency of electrons in bilayer graphene
and $m^{\ast}$ is the effective mass given as $0.044m_{e}$ with $m_{e}$ being
the bare electron mass. The Landau level spectrum of electrons given by Eq.(4)
is distinctly different from that of monolayer graphene and conventional 2DEG
system. The electrons in bilayer are quasiparticles that exhibit parabolic
dispersion with a smaller effective mass than the standard electrons. Bilayer
graphene has four fold degenerate (spin and valley) states other than the
$n=0$ level with energy $\varepsilon(n=0)=0$ which is eight-fold degenerate.
These quasiparticles are chiral exhibiting $2\pi$ Berry's phase.

\subsection{INTER-LANDAU-BAND PLASMON SPECTRUM OF MONOLAYER AND\ BILAYER
GRAPHENE IN\ A MAGNETIC FIELD}

The dynamic and static response properties of an electron system are all
embodied in the structure of the density-density correlation function. We
employ the Ehrenreich-Cohen self-consistent-field (SCF) approach \cite{22} to
calculate the density-density correlation function. The SCF treatment
presented here is by its nature a high density approximation which has been
successfully employed in the study of collective excitations in
low-dimensional systems both with and without an applied magnetic field. It
has been found that SCF predictions of plasmon spectra are in excellent
agreement with experimental results. Following the SCF approach, one can
express the dielectric function as%
\begin{equation}
\epsilon(\bar{q},\omega)=1-v_{c}(\bar{q})\Pi(\bar{q},\omega). \label{7}%
\end{equation}
where the two-dimensional Fourier transform of the Coulomb potential
$v_{c}(\bar{q})=\frac{2\pi e^{2}}{\kappa\overline{q}}$, $\overline{q}%
=(q_{x}^{2}+q_{y}^{2})^{1/2},\kappa$ is the background dielectric constant and
$\Pi(\bar{q},\omega)$ is the non-interacting density-density correlation
function
\begin{align}
\Pi(\bar{q},\omega)  &  =\frac{2}{\pi l^{2}}\sum C_{nn^{\prime}}\left(
\frac{\bar{q}^{2}}{2eB}\right)  [f(\varepsilon(n)-f(\varepsilon(n^{\prime
}))]\nonumber\\
&  \times\lbrack\varepsilon(n)-\varepsilon(n^{\prime})+\omega+i\eta]^{-1},
\label{8}%
\end{align}
where $C_{nn^{\prime}}\left(  x\right)  =(n_{2}!/n_{1}!)\left(  x\right)
^{n_{1}-n_{2}}e^{-x}\left[  L_{n_{2}}^{^{n_{1}-n_{2}}}(x)\right]  ^{2}$ with
$n_{1}=\max(n,n^{\prime}),n_{2}=\min(n,n^{\prime})$, $L_{n}^{^{l}}(x)$ an
associated Laguerre polynomial with $x=\frac{\bar{q}^{2}}{2eB}$ here. This is
a convenient form of $\Pi(\bar{q},\omega)$ that facilitates writing of the
real and imaginary parts of the correlation function. The plasmon modes are
determined from the roots of the longitudinal dispersion relation%
\begin{equation}
1-v_{c}(\bar{q})\operatorname{Re}\Pi(\bar{q},\omega)=0 \label{9}%
\end{equation}
along with the condition Im$\Pi(\bar{q},\omega)=0$ to ensure long-lived
excitations. Employing Eq.(8), Eq.(9) can be expressed as
\begin{equation}
1=\frac{2\pi e^{2}}{\kappa\bar{q}}\frac{2}{\pi l^{2}}\underset{n,n^{\prime}%
}{\sum}C_{nn^{\prime}}\left(  x\right)  (I_{1}(\omega)+I_{1}(-\omega)),
\label{10}%
\end{equation}%
\begin{equation}
I_{1}(\omega)=\left(  \frac{f(\varepsilon(n))}{\varepsilon(n)-\varepsilon
(n^{\prime})+\omega}\right)  . \label{11}%
\end{equation}
and factor of $2$ due to valley degeneracy. The plasmon modes originate from
two kinds of electronic transitions: those involving different Landau bands
(inter-Landau band plasmons) and those within a single Landau-band
(intra-Landau band plasmons). Inter-Landau band plasmons involve the local 2D
magnetoplasma mode and the Bernstein-like plasma resonances, all of which
involve excitation frequencies greater than the Landau-band separation. Since,
in this work, we are not considering Landau level broadening therefore only
the inter-Landau band plasmons will be investigated.

\smallskip We now examine the inter-Landau-band transitions. In this case $n$
$\neq n^{\prime}$and Eq.(11) yields
\begin{equation}
I_{1}(\omega)=\frac{f(\varepsilon(n))}{(\omega-\Delta)},\label{12}%
\end{equation}
where $\Delta=\left(  \varepsilon(n)-\varepsilon(n^{\prime})\right)  $ which
permits us to write the following term in Eq (10) as%
\begin{equation}
(I_{1}(\omega)+I_{1}(-\omega))=2\frac{\Delta f(\varepsilon(n))}{(\omega
)^{2}-(\Delta)^{2}}.\label{13}%
\end{equation}
Next, we consider the coefficient $C_{nn^{\prime}}(x)$ in Eq.(10) and expand
it to lowest order in its argument (low wave-number expansion). In this case,
we are only considering the $n^{\prime}=n\pm1$ terms. The inter-Landau band
plasmon modes under consideration arise from neighboring Landau bands. Hence
for $n^{\prime}=n+1$ and $x\ll$ 1, using the following associated Laguerre
polynomial expansion $L_{n}^{^{l}}(x)=%
{\displaystyle\sum\limits_{m=0}^{n}}
(-1)^{m}\frac{(n+l)!}{(l+m)!(n-m)!}\frac{x^{m}}{m!}$ for $l>0$ \cite{23} and
retaining the first term in the expansion for  $x\ll$ 1, $C_{nn^{\prime}}(x)$
reduces to
\begin{equation}
C_{n,n+1}(x)\rightarrow(n+1)x,\label{14}%
\end{equation}
and for $n^{\prime}=n-1$ and $x\ll1,$ it reduces to
\begin{equation}
C_{n,n-1}(x)\rightarrow nx.\label{15}%
\end{equation}
Substitution of equations (13) and (14, 15) into equation (10) and replacing
$x=\frac{\bar{q}^{2}}{2eB}$ yields
\begin{align}
1 &  =\frac{2\pi e^{2}}{\kappa\bar{q}}\frac{2}{\pi l^{2}}\underset{n}{\sum
}\left(  (n+1)\left(  \frac{\bar{q}^{2}}{2eB}\right)  \frac{2\left(
\frac{\omega_{g}}{2\sqrt{n}}\right)  f(\varepsilon(n))}{\left(  \omega
^{2}-\left(  \frac{\omega_{g}}{2\sqrt{n}}\right)  ^{2}\right)  }\right.
\nonumber\\
&  +\left.  n\left(  \frac{\bar{q}^{2}}{2eB}\right)  \frac{2\left(
-\frac{\omega_{g}}{2\sqrt{n}}\right)  f(\varepsilon(n))}{\left(  \omega
^{2}-\left(  \frac{\omega_{g}}{2\sqrt{n}}\right)  ^{2}\right)  }\right)
.\label{16}%
\end{align}
In obtaining the above result we note that $\Delta=\left(  \sqrt{n^{\prime}%
}-\sqrt{n}\right)  \omega_{g}$. Therefore, $\Delta=\frac{\omega_{g}}{2\sqrt
{n}}$ for $n^{\prime}=n+1,$ and $\Delta=-\frac{\omega_{g}}{2\sqrt{n}}$ for
$n^{\prime}=n-1$. We are considering the weak magnetic field case where many
Landau levels are filled. In that case, we may substitute $\sqrt{n_{F}}$ for
$\sqrt{n}$ in Equation (16). $n_{F}=\left(  \frac{\varepsilon_{F}}{\omega_{g}%
}\right)  ^{2}$ is the Landau level index corresponding to the Fermi energy
$\varepsilon_{F}.$ Equation (16) can be expressed as
\begin{equation}
\omega^{2}=\frac{2\pi e^{2}v_{F}}{\kappa}\bar{q}\left(  \underset{n}{\sum
}\frac{2eB}{\pi k_{F}}f(\varepsilon(n))\right)  .\label{17}%
\end{equation}
In terms of the 2D electron density $n_{2D}=\underset{n}{\sum}\frac{2eB}{\pi
}f(\varepsilon_{n})\ $the inter-Landau-band plasmon dispersion relation for
monolayer graphene can be expressed as
\begin{equation}
\omega^{2}=\frac{2\pi e^{2}v_{F}n_{2D}}{\kappa k_{F}}\bar{q}.\label{18}%
\end{equation}

Corresponding calculation for bilayer graphene can be carried out. The
equation that replaces Eq.(16), given above for monolayer graphene, is%
\begin{align}
1  &  =\frac{2\pi e^{2}}{\kappa\bar{q}}\frac{2}{\pi l^{2}}\underset{n}{\sum
}\left(  (n+1)(\frac{\bar{q}^{2}}{2eB})\frac{2(\omega_{b})f(\varepsilon_{n}%
)}{(\omega^{2}-(\omega_{b})^{2})}\right. \nonumber\\
&  +\left.  n\left(  \frac{\bar{q}^{2}}{2eB}\right)  \frac{2(-\omega
_{b})f(\varepsilon_{n})}{(\omega^{2}-(\omega_{b})^{2})}\right)  . \label{19}%
\end{align}
For bilayer graphene Eq.(19) can be expressed as%
\begin{equation}
1=\frac{4\pi e^{2}}{\kappa m^{\ast}}\bar{q}\frac{1}{\omega^{2}-(\omega
_{b})^{2}}\left(  \frac{m^{\ast}\omega_{b}}{\pi}\underset{n}{\sum
}f(\varepsilon_{n})\right)  \label{20}%
\end{equation}
If we define $n_{2D}=\frac{m^{\ast}\omega_{b}}{\pi}%
{\displaystyle\sum\limits_{n}}
f(\varepsilon_{n})$ and the plasma frequency as%
\begin{equation}
\omega_{p,2D}^{2}=\frac{4\pi n_{2D}e^{2}}{\kappa m^{\ast}}\bar{q}, \label{21}%
\end{equation}
then the inter-Landau-band plasmon dispersion relation for bilayer graphene
is
\begin{equation}
\omega^{2}=(\omega_{b})^{2}+\omega_{p,2D}^{2}. \label{22}%
\end{equation}

\subsection{DISCUSSION\ OF\ RESULTS}

Eqs.(18) and (22) are the central results of this work. Eq.(18) is the
inter-Landau band plasmon dispersion relation for monolayer graphene. The
inter-Landau band plasmon energy as a function of the inverse magnetic field
for the monolayer and bilayer graphene system with the plasmon energy for 2DEG
at zero temperature is presented in Figs.(1,2). The following parameters were
employed for doped graphene ($\operatorname{Si}O_{2}$ substrate): $\kappa
=2.5$, $n_{2D}=3\times10^{15}$ m$^{-2}$, $v_{F}=10^{6}$m/s. For the
conventional 2DEG (a 2DEG at the GaAs-AlGaAs heterojunction) we use the
following parameters: $m=.07m_{e}$($m_{e}$ is the electron mass), $\kappa=12$
and $n_{2D}=3\times10^{15}$ m$^{-2}.$ For electron density and magnetic field
considered, electrons fill approximately 30 Landau levels, the upper limit in
the summation for $n_{2D}$ is taken to be $n=30$ while the lower limit is
$n=0.$ In Fig.(1) we have plotted the plasmon energy as a function of the
inverse magnetic field for both monolayer graphene and conventional 2DEG. The
SdH-type oscillations are clearly visible that are a result of emptying out of
electrons from successive Landau levels when they pass through the Fermi level
as the magnetic field is increased. The amplitude of these oscillations is a
monotonic function of the magnetic field. These oscillations have a $\pi$
Berry's phase due to the chiral nature of the quasiparticles in this system,
the phase acquired by Dirac electrons in the presence of a magnetic
field\cite{1}. We also observe that the plasmon energy is $\sim4.2$ times
greater than in the 2DEG for the parameters considered. This is essentially
due to the higher Fermi energy of the electrons in graphene and the smaller
background dielectric constant.

For bilayer graphene, we consider Eq.(22). There are two main differences
between the plasmon dispersion relation for bilayer graphene given in Eq.(22)
and the standard 2DEG result. Firstly, the cyclotron frequency $\omega_{b}$ in
bilayer is $\thicksim2$ greater than the cyclotron frequency $\omega_{c}$ at
the same magnetic field in 2DEG due to the difference in the effective masses
of the electrons in the two systems. Secondly, the 2D plasma frequency
$\omega_{p,2D}$ is also larger than in 2DEG for the same wave number $\bar{q}$
due to the smaller effective mass of electrons in bilayer compared to 2DEG and
the smaller background dielectric constant $k=3$ in bilayer. The inter-Landau
band plasmon energy as a function of the inverse magnetic field for doped
bilayer and the 2DEG is shown in Fig.(2). The following parameters were used
($\operatorname{Si}O_{2}$ substrate): $\kappa=3$, $n_{2D}=3\times10^{15}$
m$^{-2}$ and $m^{\ast}=0.044m_{e}$with $m_{e}$ being the usual electron mass.
We again observe the SdH-type oscillations whose amplitude is a monotonic
function of the magnetic field. We observe that the plasmon energy is
$\sim2.6$ times greater than in the 2DEG due to the smaller effective mass,
valley degeneracy and smaller background dielectric constant. Due to the
chiral nature of the quasiparticles in bilayer graphene, $2\pi$ Berry's phase
is evident in the SdH type oscillations displayed in Fig.(2).

In conclusion, we have determined the inter-Landau band plasmon frequency for
both monolayer and bilayer graphene employing the SCF approach. The
inter-Landau band plasmon energy is presented as a function of the inverse
magnetic field. The SdH-type oscillations are clearly visible in both the
systems and their amplitude is a monotonic function of the magnetic field. Due
to the chiral nature of the quasiparticles in the mono and bilayer graphene
system, $\pi$ and $2\pi$ Berry's phases are observed in the SdH- type
oscillations in the plasmon spectrum.

One of us (K.S.) would like to acknowledge the support of the Pakistan Science
Foundation (PSF) through project No. C-QU/Phys (129). M. T. would like to
acknowledge the support of the Pakistan Higher Education Commission (HEC).

$\ast$Present address: Department of Physics, Blackett Laboratory, Imperial
College London, London SW7 2AZ, United Kingdom.

$^{1}$ Email: mtahir06@imperial.ac.uk

$^{2}$ Email: ksabeeh@qau.edu.pk, kashifsabeeh@hotmail.com.

\end{document}